%% file: ma348.tex
\documentstyle[psfig]{mn} 

\title{The shape of pulsar radio beams}

\author[J. L. Han \& R. N. Manchester]
       {
        J. L. Han$^{1,2}$,
        R. N. Manchester$^3$
\\
$^1$Beijing Astronomical Observatory of National Astronomical Observatories, 
   CAS, Beijing 100012, China; hjl@bao.ac.cn \\
$^2$Beijing Astrophysical Center, CAS-PKU, Beijing 100871, China\\
$^3$Australia Telescope National Facility, CSIRO, PO Box 76,
        Epping, NSW 2121, Australia; rmanches@atnf.csiro.au
	}

\date{Accepted \hspace{2cm};
        Received \hspace{2cm};
	in original form
     }
\pubyear{2000}
\begin{document}
\maketitle


\begin{abstract}
Using all available multi-component radio pulse profiles for pulsars
with medium to long periods and good polarisation data, we have
constructed a two-dimensional image of the mean radio beamshape. This
shows a peak near the centre of the beam but is otherwise relatively
uniform with only mild enhancements in a few regions. This result
supports the patchy-beam model for emission beams in which the mean
beam shape represents the properties of the emission mechanism and
observed pulse components result from emission sources distributed
randomly across the beam.
\end{abstract}

\begin{keywords}
Pulsars: general
\end{keywords}

\section{Introduction}
Pulsars are generally believed to be rotating neutron stars in which
the observed pulses result from one or more emission beams sweeping
across the Earth as the star rotates. Observations of radio
polarisation \cite{rc69a} led to the magnetic-pole model in which the
emission beam was centred on the magnetic axis of a predominantly
dipole magnetic field. On the assumption that the emission is directed
radially, the observed pulse profile reflects the variations in
emission intensity along a line of constant rotational
latitude. Although more than 1000 pulsars are now known
\cite{tml93,lcm+00} the shape of pulsar radio beams remains
controversial.

There are two main areas of uncertainty. One concerns the outline
shape of the radio beam. Originally assumed to be circular, some
investigations (e.g. Narayan \& Vivekanand 1983)\nocite{nv83} argued for an
elliptical beam extend in the latitude direction with a large axial
ratio. In contrast, Biggs (1990)\nocite{big90b} suggested that the beam
was slightly compressed in the latitudinal direction. Most other
investigations (e.g. Lyne \& Manchester 1988; Bj\"ornsson 1998; Gil \&
Han 1996)\nocite{lm88,bjo98,gh96} have concluded that the emission
beam is essentially circular, and we will assume this in the present
investigation.

The second area of uncertainty concerns the form of the beam pattern.
Early observations showed that there are two or more pulse components
in many pulsars (e.g. Lyne, Smith \& Graham 1971; Manchester
1971)\nocite{lsg71,man71b}.  Ruderman \& Sutherland (1975)\nocite{rs75}
presented a detailed model for pulsar radio emission in which the beam
had the shape of a hollow cone, more intense around the periphery,
corresponding to the last open field lines emanating from the polar
cap region. The radio emission was attributed to curvature radiation
by positrons moving these field lines. Such radiation is linearly
polarised in the plane of curvature of the magnetic field and
hence the model naturally explained the smooth sweep of polarisation
position angle seen in many pulsars, as well as the common occurrence
of double-peaked pulse profiles.  

Backer (1976)\nocite{bac76} first discussed the more-or-less central pulse
component seen in many pulsars and suggested that it resulted from an axial
beam.  Rankin (1983)\nocite{ran83a} made the distinction between this
central or `core' emission and the outer or `conal' emission, and showed
that these two components of the profile had rather different properties. At
least in short-period pulsars, core emission has a steeper spectrum than
conal emission \cite{ran83a,lm88} and the fluctuation properties of the two
types of emission are very different \cite{ran86}.  This model was
subsequently extended to have two or more coaxial conal emission zones,
either to account for multiple-component profiles \cite{gk97,ql98} or for
the appearance of components at different apparent radii from the conal axis
\cite{ran93,kwj+94,md99}. These latter analyses depend on a knowledge of the
angle between the beam and rotation axes ($\alpha$), usually computed from
the observed width of a `core' component. Unfortunately, the determination
of this angle is very model dependent and this results in large
uncertainties in the derived radii.

Based on a large sample of pulse shape and polarisation data, Lyne \&
Manchester (1988) found that pulse components in multiple-component
profiles were not symmetrically located within the beam
boundary. Furthermore, they were generally of very different
intensities and, in some cases, missing altogether. Their results were
consistent with components being randomly located within the beam
boundary, leading to the `patchy-beam' model. Gould
(1994)\nocite{gou94} and Gould, Lyne \& Smith (2000)\nocite{gls00}
confirmed these conclusions with an even larger data set. 

\input ma348_tab.tex

Manchester (1995)\nocite{man95b} suggested an interpretation of these
results in which the observed pulse profile is the product of a `window
function', which is a function of pulse period and radio frequency
but common to all pulsars, and a `source function' which determines
the strength of the emission at a given point within the beam and is different
for every pulsar. For example, the source function may be determined
by the density or energy of the plasma beams along different field lines,
whereas the window function is determined by the properties of the
emission mechanism.

In this paper, we determine the two-dimensional shape of the window function
at frequencies around 1 GHz by averaging the observed pulse profiles of
multi-component pulsars having reliable polarisation data. This analysis
depends only on the observed width of the pulse profile and the `impact
parameter', that is the minimum angle between the line of sight and the beam
axis, usually given the symbol $\beta$. In contrast to the beam inclination
angle $\alpha$, the impact parameter is normally well determined from the
maximum rate of change of position angle at the profile centre
\cite{lm88}. We assume that emission is seen across the entire polar cap and
choose only pulsars with profiles where this appears to be the case. In
Section 2 we describe the data set and our methods of determining the
two-dimensional beam pattern. Results are presented and discussed in Section
3 and Section 4 summarises our conclusions.

\section{The data set and analysis method}
In order to reliably determine the two-dimensional beam shape we
choose pulsars which have pulse profiles with two or more pulse
components, good signal/noise ratio and good quality polarisation
data. 

For most pulsars with two or more pulse components, we see emission from
right across the polar cap. The polarisation information allows a
measurement of the normalised impact angle, $\beta_n$, that is, the impact
parameter $\beta$ expressed as a fraction of the beam radius. This parameter
is computed assuming $\alpha = 90\degr$, but it is not very sensitive to the
actual value of $\alpha$ (Lyne \& Manchester 1988). The polarisation data
also allow a check on the assumption that we are seeing emission right to
both edges of the beam. Only those pulsars for which the maximum rate of
change of position angle is close to the pulse centre were included in the
sample. Single component pulsars are not included in the sample as it is
often difficult to reliably determine if they are predominantly core
emission (low $\beta_n$) or from the outer edge of the beam (high
$\beta_n$).

Millisecond pulsars were also excluded from the sample as they generally
have very wide and complex profiles and polarisation variations which
do not fit the simple rotating-vector model (e.g. Navarro et al. 1997,
Stairs et al. 1999)\nocite{nms+97,stc99}. It is therefore difficult or
impossible to obtain reliable values of $\beta_n$ for these
pulsars. Very short-period pulsars also commonly have wide or
interpulse profiles and are excluded from the sample for similar
reasons. 

Pulse profiles in a given pulsar usually evolve with frequency, with
some components becoming stronger and other becoming weaker or even
disappearing at higher or lower frequencies. At low frequencies, the
core component tends to dominate the profile, whereas at high
frequencies the conal emission is generally dominant. To ensure relative
uniformity in the data set, we restricted our attention to profiles at
frequencies between 600 and 1600 MHz. The results we obtain therefore
represent pulsar emission at frequencies around 1 GHz.

\begin{figure}
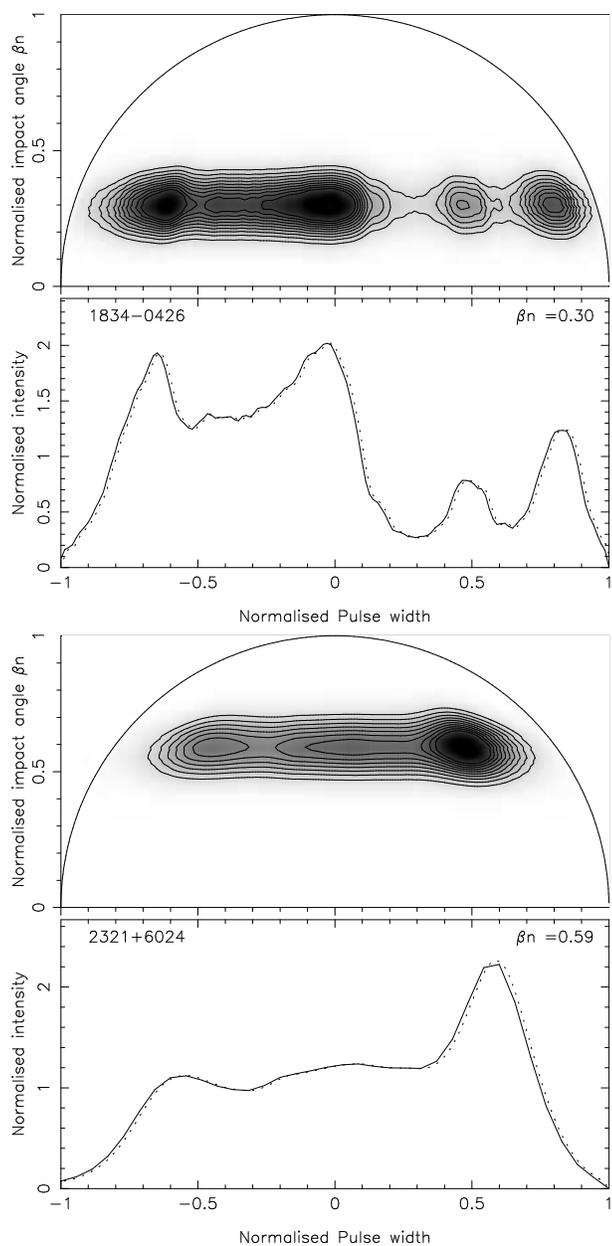
		
\begin{tabular}{c}
\mbox{\psfig{file=ma348f1a.ps,width=80mm,angle=270}} \\ 
\mbox{\psfig{file=ma348f1b.ps,width=80mm,angle=270}}
\end{tabular}
\caption[]{Beam components for four pulsars derived using the
procedures described in Section 2. The pulse profiles are given in the
lower panels where the dotted lines represent the interpolated profile
used in the projection on to the beam pattern.}
\end{figure}

We checked the profiles available on the pulsar profile database
\footnote{See http://www.mpifr-bonn.mpg.de/div/pulsar/data/} of
European Pulsar Network maintained at Max-Planck-Institut f\"ur
Radioastronomie \cite{ljs+98}, which contains pulsar profiles from
more than 50 papers, including the large datasets from Gould \& Lyne
(1998)\nocite{gl98}, Manchester, Han \& Qiao (1998)\nocite{mhq98} and
Hoensbroech \& Xilouris (1997)\nocite{hx97}.  Normalised impact
parameters were obtained from Lyne \& Manchester (1988), Rankin
(1993)\nocite{ran93b}, Gould (1994)\nocite{gou94} and Manchester et
al. (1998).  The final data set of 87 pulsars which satisfy the above
criteria is listed in Table 1.

To determine the average beamshape, we first defined the duration of
each pulse profile by taking the points at which the signal rose above
$3\sigma$, where $\sigma$ is the rms deviation of the off-pulse noise.
Next, the mean intensity of the observed emission within the pulse was
normalised to unity. The normalised impact parameter, $\beta_n$, gives
the offset of the locus of line of sight across the polar cap in the
latitude direction as a fraction of the beam radius. We assumed a
circular beam of unity radius and a straight trajectory of the
emission point across the polar cap. The beamshape is assumed to be
symmetric about the equator, i.e., the sign of $\beta_n$ is ignored.

The profile intensity was interpolated on to an $x-y$ array
representing a semi-circular beam of unity radius along a line at
$y=\beta_n$ between $x$ values of $\pm \sqrt{1-\beta_n^2}$ using a
five-point polynomial interpolation routine \cite{ptvf92}. To allow
for uncertainties in the value of $\beta_n$, possible non-linearities
in the beam trajectory across the polar cap, and to represent the
finite width of a subpulse beam, the pulse profile is broadened in
latitude $(y)$ assuming a guassian form,
$\exp[-(\beta-\beta_n)^2/0.1]$, between
$\Delta\beta=\pm0.3$. Examples of computed beam components for two
pulsars are shown in Fig. 1.

\begin{figure}		
\psfig{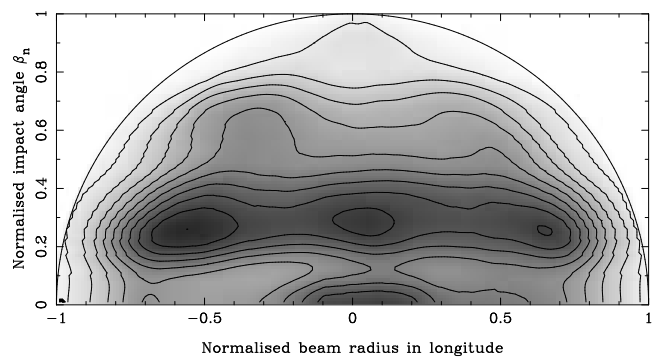}
\caption[]{The beam pattern obtained by adding data for all pulsars
in the sample represented as a greyscale. This pattern has not been
normalised for the nonuniform distribution of $\beta_n$.}
\end{figure}
\begin{figure}		
\psfig{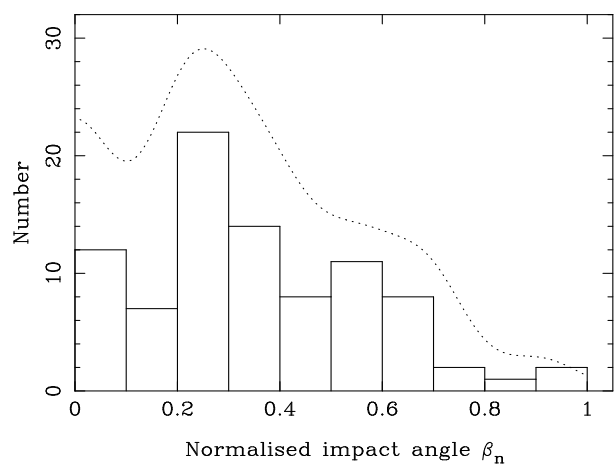}
\caption[]{Distribution of normalised impact paramters, $\beta_n$, for
pulsars in the sample. The dotted line is the sum of the
gaussian-broadened values used in the projection (see text).}
\end{figure}
\begin{figure*}		
\psfig{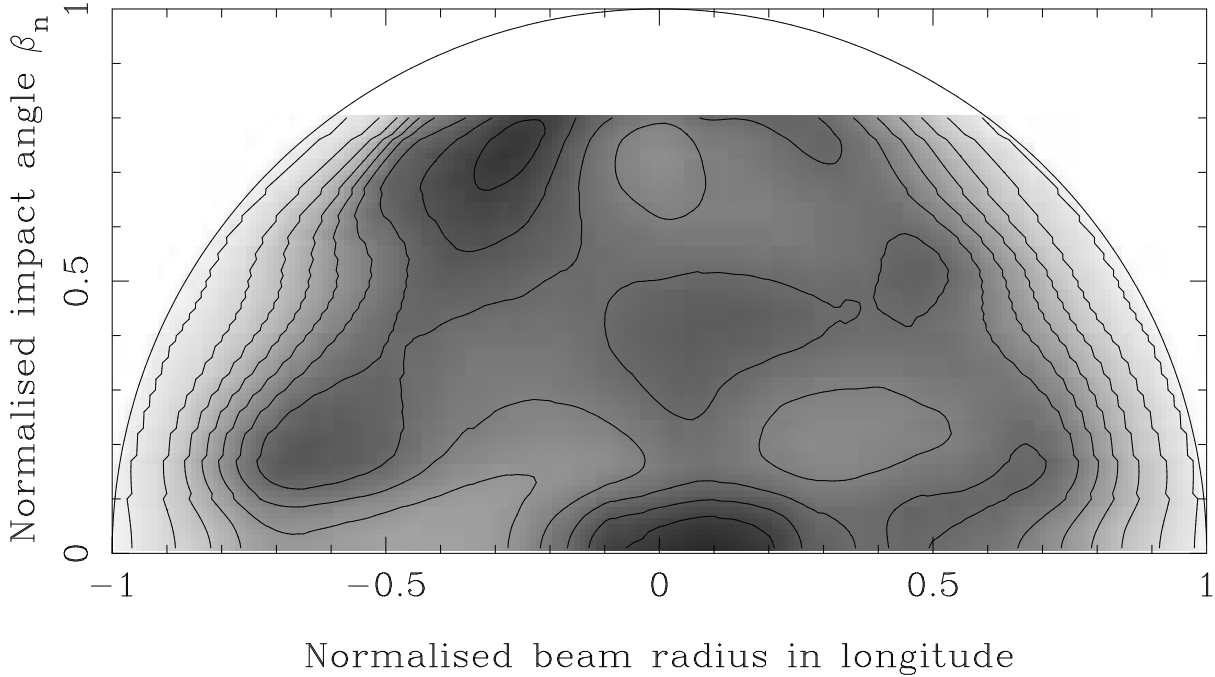}
\caption[]{Average shape of the pulsar radio beam for $\beta_n<0.8$. The
contours are at multiples of 0.1 of the peak value near the beam center. }
\end{figure*}

\section{Results and Discussion}
Integrating all pulse profiles in the sample gives the beam shape
shown in Fig. 2. This figure represents the probabilty of observing
beam components at each location. However, it is strongly affected by
the sample distribution of $\beta_n$. As shown in Fig. 3, this has a
strong peak at $\beta_n \sim 0.25$, indicating that multi-component
pulse profiles are most likely to be detected at about this impact
parameter. The deficit at $\beta_n \sim 0$ is an observational
selection effect which results from the smearing
of rapid position angle changes near the profile centre due to finite
subpulse beamwidths and instrumental broadening of the profile.
Few multi-component profiles are observed at high $\beta_n$.

The beam distribution shown in Fig. 2 was normalised to correct for
the non-uniform distribution of $\beta_n$ using the gaussian-smoothed
distribution shown in Fig. 3, giving the final average beam shape
shown in Fig. 4. Since there are only one or two pulsars in bins with
$\beta_n >0.8$ (Fig. 3), there is large statistical uncertainty in the average
profile shape for these bins. Therefore we do not plot this region of the
normalised beam pattern. 

The overall impression given by Fig. 4 is that of a `patchy'
beam. There are some systematic features, but in general the intensity
is relatively uniform across the whole beam. The central or `core'
feature is significantly displaced toward later longitudes.  The next
most prominent property of the average beam after the core component
is the presence of a few patches of slightly enhanced emission. Most of
these are located within an irregular and rather broad annular maximum
at a normalised radius of about 0.7 which is somewhat stronger on the
leading side of the beam.  Underlying this is more-or-less uniform
emission over the whole of the emission beam. The decline at the beam
edge is not very sharp.

Although there is no evidence for double or multiple cones in Fig. 4,
our present data set is dominated by two- and three-component
pulsars. A larger sample of pulsars with more than three components is
needed to reliably distinguish multiple-cone models from a patchy beam
models. However, our results suggest that, if multiple cones exist,
they are at different radii relative to the beam radius in different
pulsars. They also show that the conal emission is not confined to a
single annular region at the beam boundary.

These results are consistent with the idea that components in pulse
profiles are largely determined by the distribution of sources across
the polar cap -- the `source function'. The rather smooth distribution
distribution of intensity in the mean beam shape suggests that these
source regions are randomly distributed for different pulsars as
suggested by Lyne \& Manchester (1988) and Manchester (1995) -- the
`patchy-beam' model.  The number of identifiable source components
across the profile is mostly limited by the finite width of subpulse
beams. Except in a few cases with very high signal/noise ratio,
e.g. PSR B0740$-$28 \cite{kra94}, it is generally not possible to
identify more than four or five peaks or components, and often only
one or two. Some very wide components identified by gaussian fitting,
e.g. PSRs B2319+60, B2021+51 \cite{kra94}, are most probably regions
of distributed emission producing overlapping subpulse beams.

We believe Fig. 4 is a good representation of the {\it mean} radio beam (for
$\beta_n < 0.8$) emitted at frequencies around 1 GHz by `typical' pulsars,
that is, the pulsars of medium or long period which dominate the sample of
known radio pulsars. It represents the `window function' in the model of
Manchester (1995) which is determined by the effective gain or efficiency of
the radio emission process.

\section{Conclusions}
We have computed the average radio beamshape at frequencies about 1
GHz of pulsars with medium to long periods. This beamshape has a peak
near its centre and a mild, broad and rather irregular enhancement at
a normalised beam radius of about 0.7, but is otherwise rather
uniform.  The decline at the beam edge is gradual.

These results suggest that the presence and location of profile
components are determined by a `source function' which varies randomly
from pulsar to pulsar. The summing of these randomly distributed
components results in a relatively uniform average beam profile which
we interpret as the `window function' representing the properties of
the emission process common to all longer-period pulsars.

\section*{Acknowledgments}
We thank Prof. Qiao Guojun and Zhao Yongheng for helpful discussions
and the referee for helpful comments.
This work was initiated during JLH's visit to the ATNF during
1997. JLH thanks the Su Shu Huang Astrophysics Research Foundation of
CAS and the exchange program between CAS and CSIRO for support of
visits in 1997 and 1999, respectively. His work in China is supported
by the National Natural Science Foundation of China and
the National Key Basic Research Science Foundation.
The profile data were obtained from the pulsar database of the European
Pulsar Network at the Max-Planck-Institut f\"ur Radioastronomie.
The Australia Telescope is funded by the Commonwealth Government
for operation as a National Facility managed by CSIRO.

\end{document}

%% file: ma348_tab.tex
\begin{table*}  
\caption{Normalized impact angles and the frequency of profile samples}
\begin{tabular}{crcccrcccrcccrc}
\hline 
 PSR J& freq& $\beta_n$& &PSR J & freq& $\beta_n$& & PSR J & freq&$\beta_n$& & PSR J& freq&$\beta_n$ \\ 
\hline
J0102$+$6537&1408&0.27&& J1136$+$1551&1408&0.69&& J1807$-$0847&1408&0.25&& J1954$+$2923&1408&0.70\\ 
J0108$+$6905& 610&0.20&& J1239$+$2453&1400&0.04&& J1810$-$5338& 660&0.93&& J2002$+$4050&1408&0.55\\
J0152$-$1637& 610&0.20&& J1509$+$5531& 925&0.52&& J1816$-$2649& 606&0.59&& J2004$+$3137&1408&0.05\\
J0332$+$5434& 925&0.26&& J1559$-$4438&1502&0.39&& J1823$+$0550& 610&0.20&& J2006$-$0807&1408&0.43\\
J0406$+$6138& 610&0.21&& J1604$-$4909& 658&0.02&& J1826$-$1334&1408&0.12&& J2022$+$2854& 925&0.52\\
J0450$-$1248& 610&0.86&& J1646$-$6831& 660&0.02&& J1829$-$1751& 925&0.02&& J2037$+$3621& 606&0.36\\
J0452$-$1759&1404&0.58&& J1651$-$1709& 606&0.70&& J1834$-$0426& 606&0.30&& J2046$+$1540&1408&0.53\\
J0528$+$2200& 925&0.14&& J1651$-$5222& 658&0.54&& J1841$+$0912&1408&0.31&& J2048$-$1616& 925&0.21\\
J0536$-$7543& 663&0.26&& J1703$-$3241& 610&0.26&& J1842$-$0359&1408&0.17&& J2053$-$7200& 658&0.18\\
J0624$-$0424&1408&0.21&& J1720$-$2933&1408&0.68&& J1847$-$0402&1408&0.35&& J2055$+$2209& 606&0.71\\
J0653$+$8051&1408&0.21&& J1733$-$2228& 610&0.69&& J1848$-$0123&1642&0.40&& J2113$+$4644& 925&0.13\\
J0729$-$1836& 610&0.67&& J1735$-$0724&1408&0.02&& J1900$-$2600& 610&0.22&& J2157$+$4017& 610&0.45\\
J0754$+$3231& 610&0.18&& J1740$+$1311&1408&0.27&& J1906$+$0641&1408&0.44&& J2212$+$2933& 610&0.26\\
J0837$+$0610&1408&0.74&& J1741$-$0840& 610&0.39&& J1907$+$4002&1408&0.26&& J2229$+$6205& 610&0.41\\
J0846$-$3533&1440&0.07&& J1745$-$3040&1642&0.03&& J1912$+$2104& 610&0.20&& J2308$+$5547&1408&0.23\\
J0907$-$5157& 660&0.52&& J1748$-$1300& 610&0.28&& J1916$+$0951& 610&0.68&& J2317$+$2149&1408&0.38\\
J0908$-$1739&1408&0.57&& J1750$-$3157&1408&0.21&& J1919$+$0021& 610&0.48&& J2321$+$6024& 925&0.59\\
J0921$+$6254& 606&0.33&& J1754$+$5201&1408&0.38&& J1921$+$1948& 610&0.57&& J2324$-$6054& 660&0.40\\
J0955$-$5304& 658&0.02&& J1756$-$2435&1408&0.31&& J1921$+$2153& 925&0.64&& J2325$+$6316& 610&0.13\\
J1034$-$3224& 661&0.03&& J1757$-$2421&1408&0.06&& J1932$+$1059&1642&0.92&& J2330$-$2005&1408&0.22\\
J1036$-$4926& 658&0.34&& J1801$-$2920&1440&0.02&& J1945$-$0040& 610&0.31&& J2337$+$6151&1408&0.42\\
J1041$-$1942& 925&0.34&& J1803$-$2137&1408&0.27&& J1948$+$3540&1408&0.36&&   \\
\hline       
\end{tabular}
\end{table*}